# Quantum theory of hydrogen key of point mutation in DNA

Ivanova E.K., *Turaeva N.N., Oksengendler B.L.,

Institute of polymer chemistry and physics,

100128 Tashkent, Kadiri str. 7B, Uzbekistan

## Abstract

Quantum theory of hydrogen atoms distribution between two complementary nucleotide bases in DNA double helix at moment of replication has been proposed in this work. It bases on two mechanisms of proton tunneling: the Andreev-Meyerovich mechanism with spontaneous phonon radiation and the Kagan-Maximov (Flynn-Stoneham) mechanism at phonon scattering. According to the presented model, the probability of proton location in shallow potential well (point mutation form) is directly proportional to temperature. It was shown that the point mutation probability decreases with increasing replication velocity.

## I. Introduction

The data received on a human are the evidence of close connection between point mutation formation and cell dividing processes (1-4). A big interest to study of this problem is derived from the Watson-Crick hypotheses of point mutation formation mechanism (2). Nucleotide bases of DNA can be in various tautomer forms. In living cells, they are in their normal form and the tautomer forms realizes with small probability. Watson and Crick supposed that the point mutation could take place when nucleotide bases were in their rare tautomer forms. During the replication of DNA instead of normal combinations of complementary nucleotide bases as A-T and G-C other combinations like $A^*$-C, A-$C^*$, $G^*$-T, G-$T^*$ are possible. If nucleotide bases are in their tautomer forms at moment of replication, the sequence of bases in recovered DNA will be different (fig.1).

* Corresponding author: nturaeva@hotmail.com

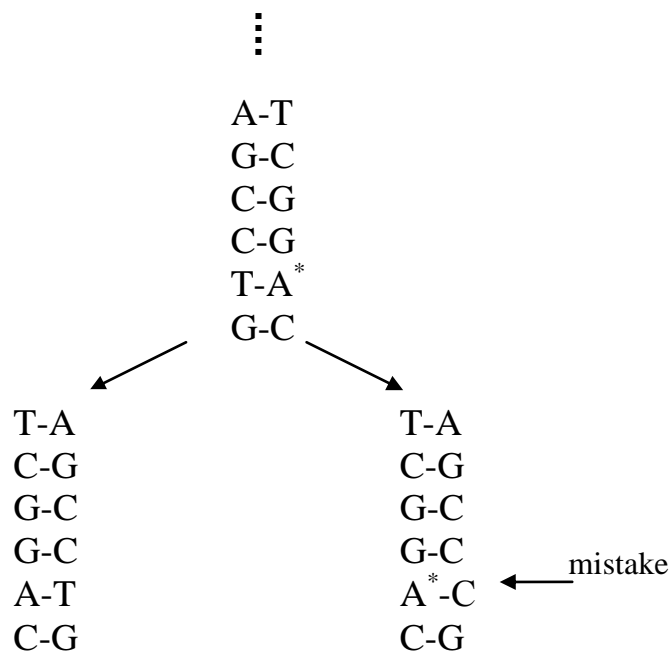

Fig. 1. Tautomer form of DNA leading to the mistake of structure

There are the Watson-Crick and the Lowdin mechanisms (4) of tautomer transition of nucleotide base pair. These transitions connect with proton tunneling along and across the DNA helix (fig. 2).

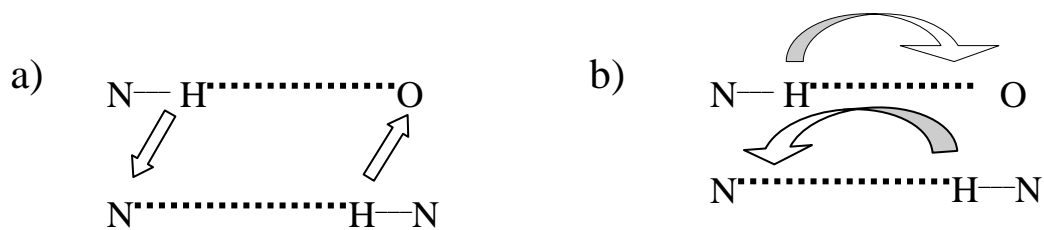

Fig. 2. Proton tunneling along (Watson-Crick mechanism) (a)

and across the DNA double helix (Lowdin mechanism) (b)

The Lowdin mechanism of proton tunneling across the DNA double helix is more probable (4). The proton transfer between two complementary nucleotide bases of DNA is described by double well potential (fig.3). However, it is known from the double well potential theory (5), that the occupation probability of each potential well has a tendency to the value of ½ with increasing temperature. Hence, increasing temperature increases the probability of tautomer form formation giving rise to spontaneous point mutations. For example at SHF irradiation of DNA increasing temperature may lead to this effect. Thus developing the quantum theory of the Lowdin mechanism of point mutations is of a great scientific interest.

## II. Quantum theory of hydrogen bond

According to the Lowdin mechanism the point mutations key is the change of proton position which connects two complementary bases (A-T and G-C) during the separation of DNA strands (replication process). During the replication proton can remain in its deep potential well 1 or transfer into the neighbor shallow well 2 forming the rare tautomer configuration (fig. 3).

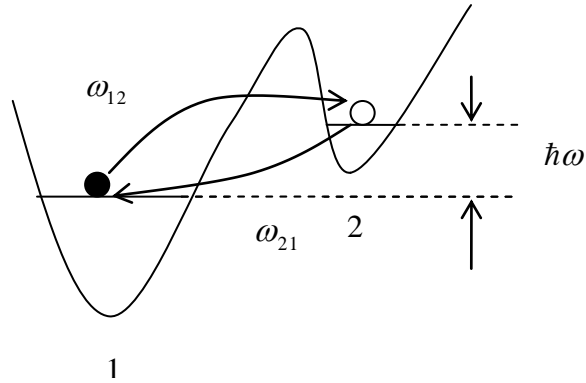

Fig. 3. The double-well potential for proton tunneling

The kinetics of proton transfer from the well 1 into the well 2 is defined by

$$\frac{dn_1}{dt} = -\omega_{12} n_1 + \omega_{21}(1-n_1). \qquad [1]$$

Here $n_1$ is the occupation number of the well 1, $\omega_{12}$ is the probability of proton transfer from the well 1 into the well 2, $\omega_{21}$ is the probability of proton transfer from the well 2 into the well 1.

In general, proton as a quantum particle can transfer through the potential barrier by two mechanisms: tunneling (quantum diffusion) (6,7,8) and thermo-activated mechanisms. The critical temperature of predominance of tunneling mechanism or thermo-activated mechanism can be evaluated by the formula proposed by Goldansky V.I. (9)

$$T_t = \frac{\sqrt{2}\hbar}{\pi k d}\sqrt{\frac{Q}{m}}. \qquad [2]$$

here k is the Boltsman constant, d is the width of potential barrier, Q is the height of potential barrier, m is the mass of proton. Taking into account the potential barrier between two minima in cytosine or guanine of 1.7 eV (10) and its width of 1A we can calculate this critical temperature, which is 1.3 times larger than the critical temperature of 320K for proton tunneling in chemical reaction with potential barrier of 1 eV. It means the principal mechanism of proton transfer in DNA at physiological temperatures is the tunneling mechanism that Lowdin intuitively supposed.

In case of different depths of potential wells, there are two mechanisms of proton tunneling. The first mechanism of proton tunneling takes place at phonon scattering on it (incoherent tunneling). The probability of this process was estimated by Kagan-Maximov (11) and defined by the following equation

$$\omega_{K-M} = \left(\frac{\Delta^2}{\hbar \theta_D}\right)\left(\frac{T}{\theta_D}\right)^7. \qquad [3]$$

Here $\Delta$ is the proton transfer integral, $\theta_D$ is the Debye temperature, $T$ is the medium temperature. The second tunneling mechanism was characterized by simultaneous spontaneous phonon radiation and proposed by Andreev and Meyerovich for quantum crystals (12). The probability of proton tunneling with simultaneous spontaneous phonon radiation is directly proportional to phonon frequency and transfer integral as follows:

$$\omega_{A-M} = \left(\frac{\Delta^2}{\hbar\theta_D}\right)\left(\frac{\hbar\omega}{\theta_D}\right)^3. \qquad [4]$$

Here $\hbar\omega$ is the lowest energy difference between the oscillation levels of neighbor potential wells. We can suppose that proton tunneling between two complementary nucleotide bases in DNA is analogous with the phenomenon of quantum diffusion in solid states. It is really that according to the theory of quantum diffusion (6-9) the object investigated is a tunneling light atom in double well potential formed by crystals. Supposing the light particle transfer takes place in the Debye phonon continuum and the Debye temperature of DNA in some medium is higher than the physiological temperatures, we can apply the Andreev-Meyerovich and Kagan-Maximov mechanisms to the process of proton transfer in DNA. Then taking into account the following

$$\begin{aligned}\omega_{12} &= \omega_{K-M} \\ \omega_{21} &= \omega_{K-M} + \omega_{A-M}\end{aligned}, \qquad [5]$$

we can write the following kinetics equation for the proton occupation number of well 1

$$\frac{dn_1}{dt} = \frac{\Delta^2}{\hbar Q_D}\left[n_1 \cdot \left[-2\left(\frac{T}{Q_D}\right)^7 - \left(\frac{\hbar\omega}{Q_D}\right)^3\right] + \left(\frac{T}{Q_D}\right)^7 + \left(\frac{\hbar\omega}{Q_D}\right)^3\right] \qquad [6]$$

For evaluation, we can take the proton transfer integral in the form of exponential dependence on its deviation from the equilibrium distance between two wells $R_o$ taking into account the Frank-Condon factor for proton oscillation ground states

$$\Delta = \Delta_0 \exp(-\gamma(R-R_0)^2), \qquad [7]$$

here $\Delta_0$ is the transfer integral at $R = R_0$, $\gamma$ is the material constant which is reversely proportional to the De Boer parameter (6) of proton in DNA.

The velocity of DNA double helix replication is different for various organisms. Usually this rate is significantly less than the proton tunneling time: $R_o\omega_{ij} >> \upsilon$. Taking into account the velocity of DNA double helix replication, we can write the formula:

$$(R-R_o)^2 = \chi\upsilon^2 t^2. \qquad [8]$$

Here $\chi$ takes into account the geometry connected with the distance between wells at different moments of replication.

In result of this substitution, one can obtain the following kinetic equation for the occupation number of well 1

$$\frac{dn_1}{dt} = \frac{\Delta_0^2 \cdot \exp(-2\gamma\chi v^2 t^2)}{\hbar Q_D} \cdot \left[ n_1 \cdot \left[ -2\left(\frac{T}{Q_D}\right)^7 - \left(\frac{\hbar\omega}{Q_D}\right)^3 \right] + \left(\frac{T}{Q_D}\right)^7 + \left(\frac{\hbar\omega}{Q_D}\right)^3 \right]. \qquad [9]$$

Solving the equation [9] at initial conditions as ($t_o = 0$, $n_1 = 1$), we receive the following expression for proton occupation number of well 1

$$n_1 = \frac{3\left(\frac{T}{Q_D}\right)^7 + 2\left(\frac{\hbar\omega}{Q_D}\right)^3}{2\left(\frac{T}{Q_D}\right)^7 + \left(\frac{\hbar\omega}{Q_D}\right)^3} \cdot \exp\left( \frac{-\Delta_0^2 \left[ 2\left(\frac{T}{Q_D}\right)^7 + \left(\frac{\hbar\omega}{Q_D}\right)^3 \right] \sqrt{\pi}}{2\hbar Q_D \sqrt{2\gamma\chi} \cdot v} \cdot \mathrm{erf}(\sqrt{2\gamma\chi} \cdot v \cdot t) \right) + \frac{\left(\frac{T}{Q_D}\right)^7 + \left(\frac{\hbar\omega}{Q_D}\right)^3}{2\left(\frac{T}{Q_D}\right)^7 + \left(\frac{\hbar\omega}{Q_D}\right)^3}$$

[10]

*Erf* function has two asymptotic values (at small and large arguments):

$$\mathrm{erf}(\sqrt{2\gamma\chi} \cdot v \cdot t) = \begin{cases} t < 1 & \dfrac{2\sqrt{2\gamma\chi} \cdot v \cdot t}{\sqrt{\pi}} \left( 1 - \dfrac{(\sqrt{2\gamma\chi} \cdot v \cdot t)^2}{3} + \dfrac{(\sqrt{2\gamma\chi} \cdot v \cdot t)^4}{10} \right) \\ t > 1 & 1 - \dfrac{\exp(-2\gamma\chi v^2 t^2)}{\sqrt{2\gamma\chi\pi} \cdot v \cdot t} + \dfrac{\exp(-2\gamma\chi v^2 t^2)}{2\gamma\chi \cdot v^2 \cdot t^2 \sqrt{\pi}} \end{cases}. \qquad [11]$$

At large time, we have the expression of

$$n_1 = \frac{3\left(\frac{T}{Q_D}\right)^7 + 2\left(\frac{\hbar\omega}{Q_D}\right)^3}{2\left(\frac{T}{Q_D}\right)^7 + \left(\frac{\hbar\omega}{Q_D}\right)^3} \cdot \exp\left( \frac{-\Delta_0^2 \left[ 2\left(\frac{T}{Q_D}\right)^7 + \left(\frac{\hbar\omega}{Q_D}\right)^3 \right] \sqrt{\pi}}{2\hbar Q_D \sqrt{2\gamma\chi} \cdot v} \left( 1 - \frac{\exp(-2\gamma\chi v^2 t^2)}{\sqrt{2\gamma\chi} \cdot v \cdot t} + \frac{\exp(-2\gamma\chi v^2 t^2)}{2\gamma\chi \cdot v^2 \cdot t^2 \sqrt{\pi}} \right) \right) + \frac{\left(\frac{T}{Q_D}\right)^7 + \left(\frac{\hbar\omega}{Q_D}\right)^3}{2\left(\frac{T}{Q_D}\right)^7 + \left(\frac{\hbar\omega}{Q_D}\right)^3}. \quad [12]$$

From the expression [12], one can see that the proton occupation number of well 1 depends on temperature and velocity of replication. Let us to analyze these dependences.

At the tendency of replication velocity to zero the proton occupation number of well 2 giving rise to spontaneous point mutation form is equal to the expression

$$1 - n_1 = 1 - \frac{\left(\frac{T}{Q_D}\right)^7 + \left(\frac{\hbar\omega}{Q_D}\right)^3}{2\left(\frac{T}{Q_D}\right)^7 + \left(\frac{\hbar\omega}{Q_D}\right)^3} = const \quad. \qquad [13]$$

At large velocities of replication, the probability of point mutations is equal to the 0:

$$1 - n_1 = 0. \qquad [14]$$

Thus at increasing the velocity of replication from zero to large values the probability of point mutations is decreased from constant to zero (fig. 4). This result is in agreement with the experimental results (13) that for bacteria with replication velocity of 1000 nucleotides per second the probability of spontaneous mutation is in the range of $10^{-8}$-$10^{-9}$, while for human DNA with replication velocity of 50 nucleotides per second it value is of $10^{-2}$- $10^{-3}$.

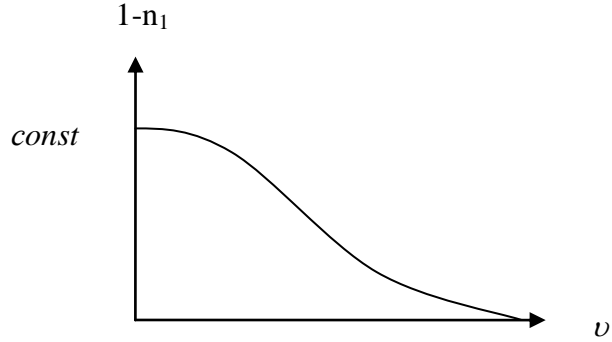

Fig. 4. Dependence of probability of spontaneous mutation on replication velocity

At low temperatures ($T << \hbar\omega$) the probability of point mutations tends to the following

$$1 - n_1 = 2 \cdot \exp\left(\frac{-\Delta_0^2 \sqrt{\pi}\left(\frac{\hbar\omega}{Q_D}\right)^3}{2\hbar Q_D \sqrt{2\gamma\chi} \cdot v}\left(1 - \frac{\exp(-2\gamma\chi v^2 t^2)}{\sqrt{2\gamma\chi\pi} \cdot v \cdot t} + \frac{\exp(-2\gamma\chi v^2 t^2)}{2\gamma\chi \cdot v^2 \cdot t^2 \sqrt{\pi}}\right)\right) = const' \qquad [15]$$

At temperatures larger than the phonon energy ($T >> \hbar\omega$) we can receive the expression of

$$1 - n_1 = \frac{1}{2}. \qquad [16]$$

From the expressions [15] and [16] it is seen that the dependence of point mutations probability on temperature can be described by the following curve smoothly growing from constant to the value of ½ (fig.5). These results are in agreement with the results which favored the view that the high temperatures increased the mutation rate (14).

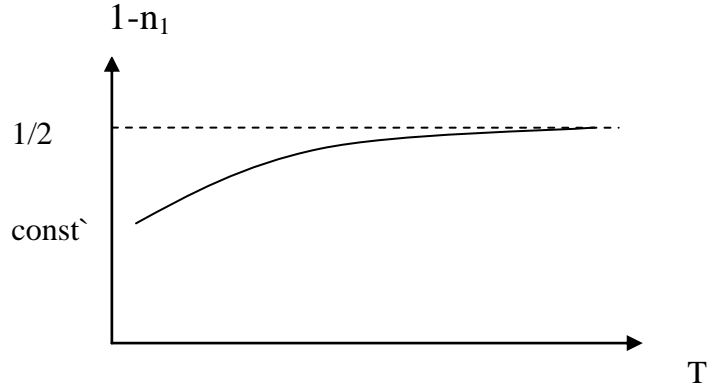

Fig. 5. Dependence of probability mutation on temperature

On the base of results of the quantum chemical and quantum dynamics calculations it was shown that the fraction of the rare tautomer of isolated cytosine and guanine are in good agreement with observed values of spontaneous GC—AT point mutations of $10^{-6}$-$10^{-8}$ before the "proof-reading" step (10). In our work, these results can be comparable with the value of $(1-n_1)/n_1$. Taking into account the maximum value of $(1-n_1)/n_1$ at low rate of replication and experimental value of GC — AT point mutations, we can evaluate the correlation of the Debye temperature of DNA in medium and physiological temperatures. From the expression of

$$\frac{1-n_1}{n_1} = \frac{\left(\frac{T}{Q_D}\right)^7}{\left(\frac{T}{Q_D}\right)^7 + \left(\frac{\hbar\omega}{Q_D}\right)^3} \approx 10^{-7} \qquad [17]$$

we can receive the following evaluation $Q_D \approx (10T)^{7/4}/(\hbar\omega)^{3/4}$. Hence, we can conclude that at any value of $\hbar\omega$ the Debye temperature of DNA system is larger than the physiological temperatures T ($Q_D > T$).

In this work we discuss the single proton transfer whereas in base pairs the double proton transfer is more probable, which is characterized by 4-well potential, as it has been initially outlined by Lowdin (4). There is known that in presence of water double proton transfer occurs via step-wise non-concerted mechanism with certain degree of proton correlation (15). However basing on the single theory of tunneling we can estimate the probability of double concerted tunneling with introduction of effective potential barrier ideas of tunneling taking into account proton-proton repulsion. The role of water molecules can be taken into account on the base of U-negative Hubbard energy (16).

### III. Conclusion

On the base of tunneling mechanisms proposed by Andreev-Meyerovich for quantum crystals and Kagan-Maximov we can analyze the probability of proton tunneling between complementary bases in DNA double helix. The probability of proton localization in shallow well during the double helix replication giving rise to spontaneous point mutations depends on the medium temperature and DNA replication velocity.

**Acknowledgement**

We are grateful to Professor A. M. Stoneham for useful discussion of application of quantum diffusion approach to the spontaneous mutation process in DNA which gives an qualitative agreement of our theory with experimental results